\documentclass[aps,prb,preprint,groupedaddress,showpacs]{revtex4}
\usepackage{graphicx}

\begin{document}


\title{Raman scattering studies of spin, charge, and lattice dynamics in
Ca$_{2-x}$Sr$_x$RuO$_4$ (0\,$\leq$\,x\,$<$\,0.2)}



\author{H. Rho,$^{1,2}$ S. L. Cooper,$^1$
S. Nakatsuji,$^{3}$ H. Fukazawa,$^3$ and Y. Maeno,$^{3,4}$ }

\affiliation{
$^{1}$Department of Physics and Frederick Seitz Materials Research Laboratory,
University of Illinois at Urbana-Champaign, Urbana, Illinois 61801\\
$^{2}$Department of Physics, Chonbuk National University, Chonju 561-756, Korea\\
$^{3}$Department of Physics, Kyoto University, Kyoto 606-8502,
Japan\\
$^{4}$International Innovation Center, Kyoto University, Kyoto
606-8501, Japan
 }


\date{\today}

\begin{abstract}

We use Raman scattering to study spin, charge, and lattice
dynamics in various phases of Ca$_{2-x}$Sr$_x$RuO$_4$.  With
increasing substitution of Ca by Sr in the range
0\,$\leq$\,x\,$<$\,0.2, we observe (1) evidence for an increase of
the electron-phonon interaction strength, (2) an increased
temperature-dependence of the two-magnon energy and linewidth in
the antiferromagnetic insulating phase, and (3) evidence for
charge gap development, and hysteresis associated with the
structural phase change, both of which are indicative of a
first-order metal-insulator transition (T$_{MI}$) and a
coexistence of metallic and insulating components for
T\,$<$\,T$_{MI}$.

\end{abstract}

\pacs{71.30.+h, 75.30.-m,75.50.Ee,78.30.-j}

\maketitle


Single-layered ruthenates have drawn much experimental and
theoretical attention due to both their structural similarity to
the high-T$_c$ cuprates, and their strongly correlated magnetic,
electronic, phononic, and orbital degrees of freedom, which result
in a rich phase diagram and exotic
phenomena.\cite{Maeno,Anisimov,Hotta,Lee,Nakatsuji1,Nakatsuji2,Braden,Nakatsuji3,Fukazawa,Alexander}
Ca$_{2-x}$Sr$_{x}$RuO$_{4}$ (CSRO) exhibits various ground states
with increasing substitution of Sr for Ca, ranging from
antiferromagnetic (AF) insulating for x = 0 to superconducting for
x = 2.
\cite{Maeno,Anisimov,Hotta,Lee,Nakatsuji1,Nakatsuji2,Braden,Nakatsuji3,Fukazawa,Alexander,Nakatsuji4}
For example, Sr$_2$RuO$_4$ (T$_c$ = 1.5 K) is a spin-triplet
superconductor which is isostructural to high-T$_c$ cuprates such
as La$_{2-x}$Ba$_2$CuO$_4$ (T$_{c}$ = 30 K).\cite{Maeno} Unlike
the doping driven high-T$_{c}$ cuprates, however, the CSRO system
is a bandwidth driven system: an increase of Ca content
significantly distorts the RuO$_6$-octahedra and decreases the
4d-band width, W, relative to the large effective Coulomb energy,
U.  As a result, the ground state for x\,$<$\,0.2 is a Mott-like
AF insulator.\cite{Nakatsuji1,Nakatsuji2,Braden,Nakatsuji3} With
increasing temperature, CSRO for x \,$<$\, 0.2 undergoes a
metal-insulator (MI) transition.  For instance, in Ca$_2$RuO$_4$,
AF ordering occurs below T$_{N}$ = 113 K, and a MI transition
occurs at T$_{MI}$ = 357 K.\cite{Fukazawa, Alexander} Slight
substitution of Ca by Sr (0 \,$<$\, x \,$<$\, 0.2) leads to a
change of both T$_N$ and T$_{MI}$. The MI transition is
first-order and is accompanied by a simultaneous structural change
with thermal hysteresis.  In this Ca-rich region, Lee et al. have
studied the temperature-dependent evolution of the optical
conductivity, finding an important role of the electron-phonon
interaction in governing the orbital arrangements, and suggesting
a possible coexistence of antiferro-orbital and the ferro-orbital
ordering states in the insulating phase.\cite{Lee} In addition,
Nakatsuji et al. recently suggested that CSRO exhibits cooperative
orbital ordering through the MI transition in the region
0\,$\leq$\,x\,$<$\,0.2.\cite{Nakatsuji3}

All of these results suggest that there are very strong
correlations among the spin, charge, lattice, and orbital degrees
of freedom in CSRO for x\,$<$\,0.2.  In this Communication, we use
the unique ability of Raman scattering to explore simultaneously
magnetic, electronic, and structural dynamics, in order to
investigate the interplay of the spin , charge , and lattice
degrees of freedom in CSRO (0\,$\leq$\,x\,$<$\,0.2) through the AF
ordering and MI transitions.

All single crystal samples, grown by a floating zone
technique,\cite{Nakatsuji1, Fukazawa,Nakatsuji4} were mounted
inside a continuous He-flow cryostat. B$_{1g}$ symmetry Raman
spectra in the crossed polarization configuration were obtained
using a Kr-ion laser with the 647.1 nm excitation wavelength in a
backscattering geometry along the c-axis. Scattered light was
dispersed through a triple spectrometer, and recorded using a
liquid-nitrogen-cooled CCD detector. All the spectra were
corrected, first, by removing the CCD dark current response, and
then by normalizing the spectrometer response using a calibrated
white light source.

\begin{figure}
\centering
\includegraphics[width=9cm]{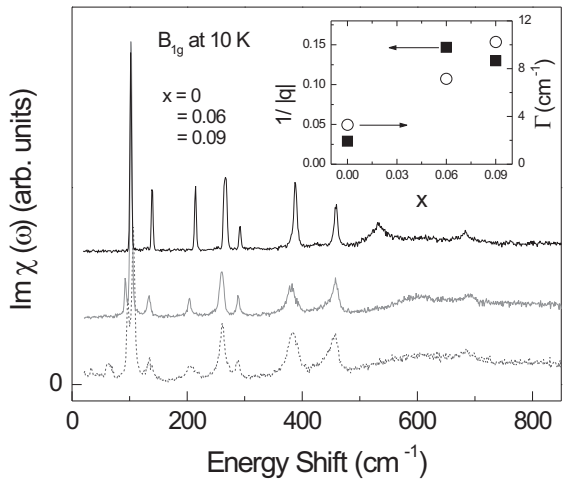}
\vspace{-0.7cm} \caption{\label{}Raman scattering spectra of CSRO
for x = 0, 0.06, and 0.09 from top to bottom, respectively.
Spectra for x $>$ 0 are shifted vertically for clarity. The inset
shows the asymmetry parameter (filled squares) and linewidth (open
circles) changes as a function of x for the 456 cm$^{-1}$ B$_{1g}$
phonon mode.}
\end{figure}

Figure\,1 shows the T\,=\,10 K Raman scattering spectra from CSRO
for x = 0, 0.06, and 0.09.  There are several key features and
trends apparent in the spectra: (i) numerous phonon lines
superimposed on the broad electronic continuum, (ii) a broad
scattering response near 600 cm$^{-1}$ associated with electronic
scattering that has developed due to the formation of a charge
gap, and (iii) a two-magnon (2M) excitation near 100 cm$^{-1}$
associated with spin-pair excitations of the Ru ions. In the
following, we consider in greater detail the temperature and x
dependence of each of these excitations.

Unlike Sr$_2$RuO$_4$, which has the ideal tetragonal K$_2$NiF$_4$
structure, Ca-substitution significantly distorts the crystal
structure of CSRO, causing both a rotation of the
RuO$_6$-octahedra around the c-axis, and a tilt of the octahedra
around an axis lying in the RuO$_2$ planes.  As a result,
Ca$_2$RuO$_4$ is orthorhombic (Pbca-D$_{2h}$$^{15}$),\cite{Braden}
and a factor group analysis indicates that there should be 9
B$_{1g}$ symmetry Raman-active optical phonons \cite{High}
involving Ca/Sr, apical oxygen, and in-plane oxygen ions.  One
notes from Fig.\,1 that while the optical phonon spectrum of
Ca$_2$RuO$_4$ exhibits narrow and symmetric phonon lineshapes at
10 K, with increasing Sr substitution there is a significant
broadening of the linewidths --- and the appearance of a distinct
``asymmetric" Fano profile --- associated with many of the
phonons. This is indicative of an increase in electron-phonon
coupling with increasing Sr substitution.\cite{Fano}  The phonon
linewidth and asymmetric parameters can be obtained from fits to a
Fano profile,
I($\omega$)=I$_0$($q$+$\epsilon$)$^2$/(1+$\epsilon$$^2$), where
$\epsilon$ = ($\omega$-$\omega$$_0$)/$\Gamma$, $\omega$$_0$ is the
phonon frequency, $\Gamma$ is the effective phonon linewidth, and
$q$ is the asymmetry parameter that is related to the
electron-phonon coupling strength V and the imaginary part of the
electronic susceptibility $\rho$ according to
$1/q$\,$\sim$\,$V\rho$.\cite{Fano,Naler} Indeed, as shown in the
inset of Fig.\,1, Sr substitution results in a significant
increase in the electron-phonon coupling and phonon linewidth
associated with the 456 cm$^{-1}$ B$_{1g}$ mode.

\begin{figure}
\centering
\includegraphics[width=9cm]{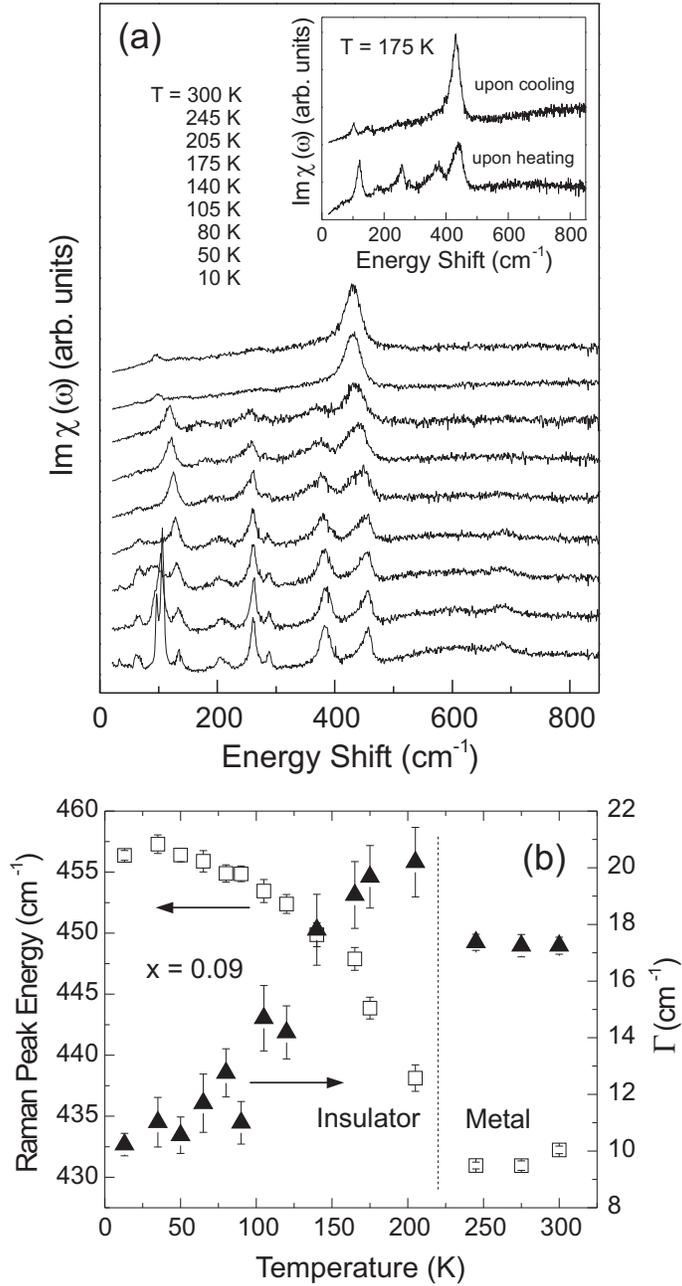}
\vspace{-0.3cm} \caption{\label{}(a) Temperature-evolution of the
Raman scattering spectra through T$_{MI}$ for the x\,=\,0.09
sample upon heating. The inset shows 175 K Raman scattering
spectra obtained by alternately cooling and heating the x\,=\,0.09
sample, showing structural hysteresis. Each spectrum is shifted
vertically for clarity. (b) Temperature-evolution of the 456
cm$^{-1}$ Raman peak energies (open squares) and linewidths
(filled triangles).}
\end{figure}

The MI transition in CSRO is driven by an elongation of the
RuO$_6$ octahedron in the metal phase.\cite{Braden} Figure\,2
illustrates that striking changes in the B$_{1g}$ Raman scattering
response of CSRO for x = 0.09 are observed through T$_{MI}$
($\sim$ 220 K) upon heating. For example, a number of phonon modes
disappear rapidly above T$_{MI}$, providing a clear demonstration
that the MI transition is accompanied by a structural change from
the S-Pbca phase to the less distorted L-Pbca phase.\cite{Braden}
Additionally, as temperature is increased toward T$_{MI}$, the
phonons soften and broaden significantly, reflecting elongation of
the RuO$_6$ octahedra along the c-axis and increased interaction
with thermally-excited carriers, respectively. These changes are
summarized in Fig.\,2 (b) for the 456 cm$^{-1}$ B$_{1g}$ mode,
which we tentatively attribute to the apical oxygen vibration,
based upon both the high frequency of this mode and a comparison
with similar modes in cuprates and other ruthenates. \cite{Sakita}
The phonon frequency and linewidth parameters were obtained at
different temperatures from fits to a Fano profile. Note that the
phonon linewidth is enhanced as the temperature approaches
T$_{MI}$, indicating a mixed phase regime and/or enhanced damping
of the phonon response due to increased structural instability
near the MI transition. Above T$_{MI}$, most of the B$_{1g}$
phonons observed below T$_{MI}$ are absent, and the two remaining
optical phonons show no further phonon frequency and linewidth
changes at least up to the room temperature, suggesting that in
the metal phase, the crystal structure is stabilized in the L-Pbca
configuration.

That the MI transition is first-order is apparent by comparing the
Raman spectra of CSRO (x = 0.09) obtained during thermal cycle.
The inset of Fig. 2 (a) illustrates that the 175 K spectrum
obtained upon cooling exhibits the simple phonon spectrum
indicative of the metallic phase, while the 175 K spectrum
obtained upon heating exhibits the complex phonon spectra
associated with the insulating phase.  This is consistent with
evidence for hysteresis observed in CSRO (0\,$\leq$\,x\,$<$\,0.2)
by transport and neutron diffraction,\cite{Braden,Nakatsuji3}
based upon which T$_{MI}$\,=\,155\,K upon cooling and 220\,K upon
heating have been estimated in CSRO for x = 0.09.

\begin{figure}
\centering
\includegraphics[width=9cm]{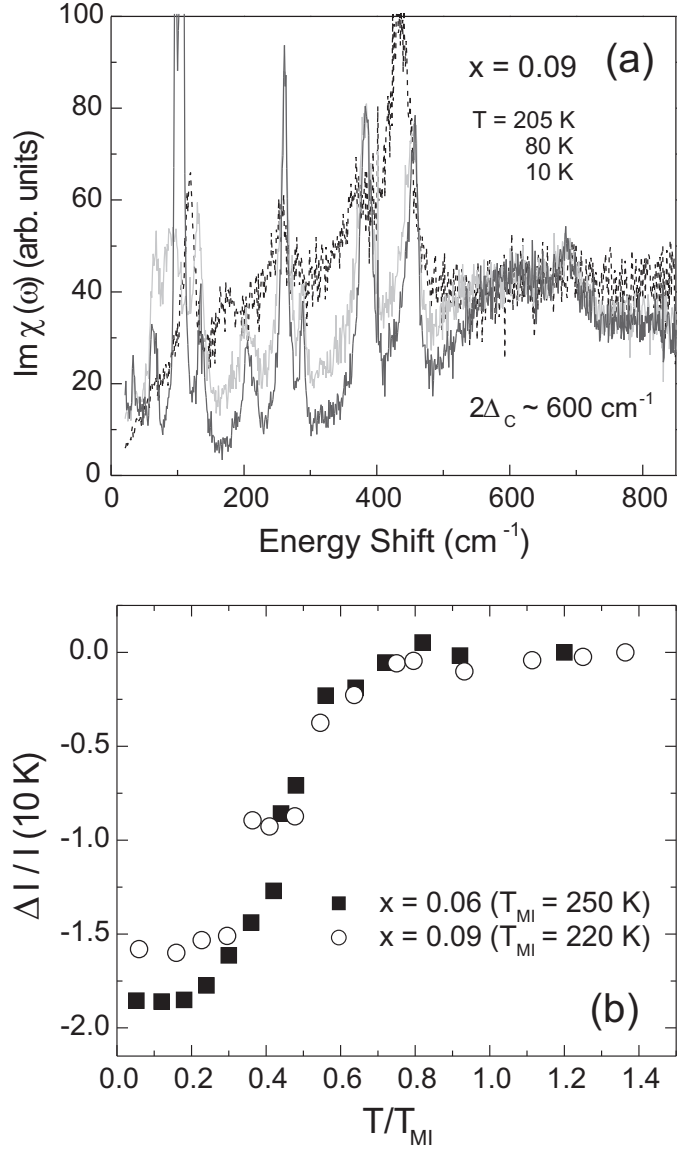}
\vspace{-0.3cm} \caption{\label{}(a) Electronic Raman scattering
spectra from the x = 0.09 sample upon heating, showing a charge
gap closing through the MI transition.  (b) Fractional changes of
the charge gap responses as a function of the reduced temperature
for x = 0.06 (filled squares) and 0.09 (open circles).}
\end{figure}

Evidence for the first-order nature of the phase transition is
also indicated in the evolution of a charge gap in the electronic
continuum below T$_{MI}$. This is illustrated, for example, in the
temperature-dependent Raman spectra of CSRO for x = 0.09 shown in
Fig.\,3 (a), in which one observes below T$_{MI}$ a suppression of
electronic scattering strength below $\sim$ 600 cm$^{-1}$, and a
redistribution of this scattering strength to a broad scattering
response between 550 and 650 cm$^{-1}$. With increasing
temperature toward T$_{MI}$, the charge gap closes systematically.
This behavior is indicative of the opening of a charge gap in
CSRO, and is similar to that observed below the MI transitions of
other correlation gap insulators such as FeSi,\cite{Nyhus1}
SmB$_6$,\cite{Nyhus2} and Ca$_3$Ru$_2$O$_7$,\cite{Liu} and
described theoretically.\cite{Freericks}  To examine the
development of the charge gap more quantitatively, we plot in
Fig.\,3 (b) the temperature-dependence of the fractional change in
the integrated electronic scattering response, $\Delta$I/I(10 K),
for x = 0.09 and 0.06, where $\Delta$I\,=\,I(T)\,-\,I(300 K) and
I(T)\,=\,\mbox{$\int_{0}^{2\Delta_c}\,Im\chi(\omega)\,d\omega$} is
the integrated spectral weight of the electronic Raman scattering
response below 2$\Delta_c$\,=\,600 cm$^{-1}$. The overall
temperature-dependence of the fractional intensity shown in
Fig.\,3 (b) is similar to those observed in FeSi, SmB$_6$, and
Ca$_3$Ru$_2$O$_7$.  There is one notable difference, however, in
that the charge gap in CSRO closes with increasing temperature
well below the MI transition temperature.  We attribute this
behavior to the likely coexistence of insulating (S-Pbca) and
metallic (L-Pbca) phase regions.  In particular, in the
temperature range between 155 K -- 220 K, CSRO at x = 0.09 is
primarily in the metal phase upon cooling, but is primarily in the
insulating phase upon heating. Consequently, during the thermal
cycle, some metal components are known to coexist with insulating
components in this temperature range. Indeed, neutron scattering
\cite{Braden}, resistivity and susceptibility \cite{Nakatsuji3}
measurements of CSRO for x $<$ 0.2 clearly indicate thermal
hysteresis through the MI transition; further, neutron scattering
measurements clearly reveal the coexistence of S-Pbca and L-Pbca
phases\cite{Braden} in this temperature range.

\begin{figure}
\centering
\includegraphics[width=9cm]{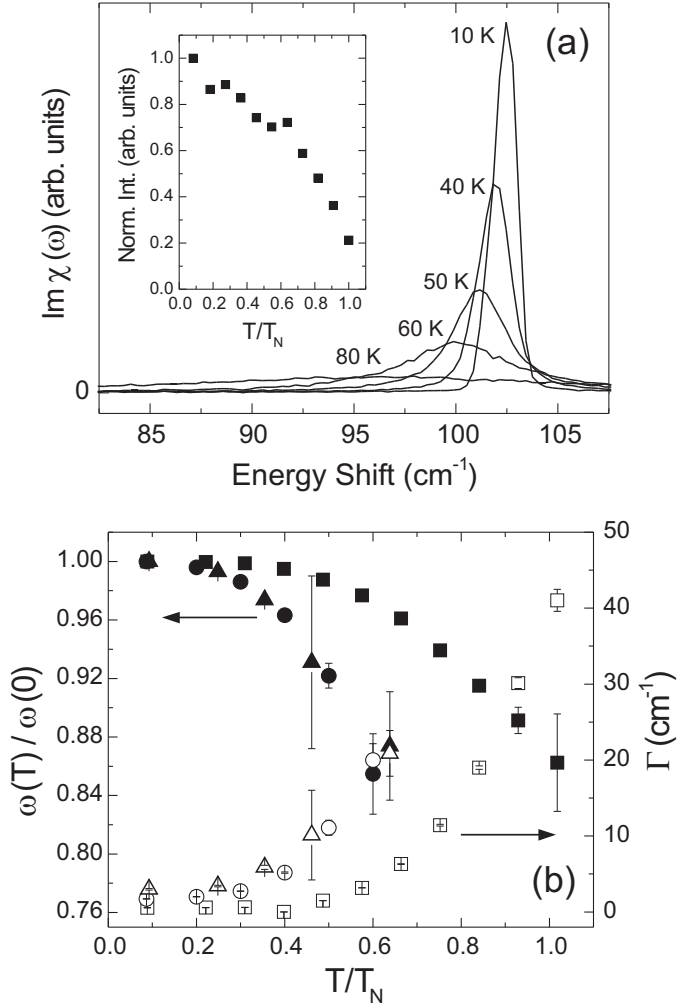}
\vspace{-0.3cm} \caption{\label{}(a) The 2M scattering response of
Ca$_2$RuO$_4$ at the elevated temperature.  The inset shows the
spectrally integrated intensity change of the 2M responses as a
function of the normalized temperature. (b) Summary of the
normalized 2M peak energy changes for x = 0 (filled squares), 0.06
(filled circles), and 0.09 (filled triangles), respectively, and
the corresponding spectral width changes for x = 0 (open squares),
0.06 (open circles), and 0.09 (open triangles), respectively, as a
function of the normalized temperature.}
\end{figure}

Finally, for T $<$ T$_N$ in CSRO at x = 0, 0.06, and 0.09, a peak
in the Raman scattering response is observed to develop rapidly
near 100 cm$^{-1}$ in the B$_{1g}$ scattering geometry (see
Figs.\,1 and 4 (a)). This response is associated with 2M
scattering, involving a photon-induced flipping of neighboring
spins on nearest-neighbor Ru-4d$^4$ sites; this scattering
response not only reveals the presence of AF correlations in the
ground state of this system, but provides useful information on
the superexchange coupling constant J.\cite{Liu, Fleury, Lyons,
Sugai, Weber} Figure\,4 summarizes the temperature-dependent
energy renormalization, linewidths, and intensity change of the 2M
scattering response of CSRO. With increasing temperature, the 2M
scattering response diminishes in intensity, shifts to lower
energies, and broadens, similar to the behavior observed in the
bilayer ruthenate system Ca$_3$Ru$_2$O$_7$.\cite{Liu} Using the
fact that the 2M peak energy for S\,=\,1 AF insulators is given by
6.7\,J,\cite{Sugai} the estimated in-plane exchange energies
between nearest-neighbor Ru-4d$^4$ sites are 15.22, 15.37, and
15.97 cm$^{-1}$ for the x\,=\,0, 0.06, and 0.09 samples,
respectively.  Interestingly, in spite of a significant change in
the N\'{e}el temperature with Sr substitution (T$_N$\,=\,113 K,
150 K, 141 K for x = 0, 0.06, and 0.09, respectively), the 2M
energy is relatively insensitive to Sr substitution, indicating
that the local AF exchange coupling is relatively unaffected by Sr
substitution at these values of x. This is consistent with recent
pressure-dependent Raman studies of Ca$_2$RuO$_4$, which show that
the 2M energies are relatively insensitive to pressure up to the
pressure-induced MI transition.\cite{Snow}  Both of these results
support the conclusion that antiferromagnetism is suppressed with
increasing Sr substitution and pressure, not by affecting the AF
exchange coupling, but rather by reducing the volume-fraction of
distorted S-Pbca phase regions that support orbital polarization
and AF order.\cite{Snow} Figure\,4 (b) also shows that the 2M
scattering responses in the x = 0, 0.06, and 0.09 samples exhibit
substantially different temperature-dependences.  In particular,
in comparison to the x\,=\,0 sample, Sr substitution causes a much
more dramatic renormalization of the 2M energy and linewidth with
increasing temperature.  This observation suggests that the Sr
substitution causes a substantially larger suppression of magnetic
correlations with increasing temperature than in the x\,=\,0
material.  This is presumably due either to the disorder
introduced in the magnetic lattice when Sr is introduced, and/or
to the increased electronic contribution introduced with Sr
substitution, which increases the amount of magnon energy and
lifetime renormalization compared to the x = 0
material.\cite{Weber}

In conclusion, Raman scattering results on CSRO for
0\,$\leq$\,x\,$<$\,0.2 provide substantial insight into the
interplay among the spin, charge, and lattice dynamics through the
MI and magnetic phase changes.  In particular, this study
demonstrates that the Sr substitution for Ca significantly affects
the magnetic and electronic excitations, as evidenced by a
substantial increase in the renormalization of the 2M energies and
linewidths, and by an increase of the electron-phonon
interactions.  The temperature-dependent evolution of the Raman
scattering response shows that the MI transition is accompanied by
both a significant change in the phonon spectrum and the
development of a charge gap. Both the hysteretic behavior of the
phonon temperature-dependence, as well as the
temperature-dependent evolution of the charge gap, are indicative
of a phase coexistence regime involving L-Pbca and S-Pbca
components near T$_{MI}$.

This work was supported by the Department of Energy through grant
DEFG02-96ER45439, and by the National Science Foundation through
grant NSF DMR-0244502. H.R. acknowledges a support of this work in
part by the research fund of Chonbuk National University.



\bibliography{basename of .bib file}

\begin{thebibliography}{}
\bibitem{High} High-resolution spectra reveal that the unresolved B$_{1g}$ phonon response at $\sim$267
cm$^{-1}$ from the Ca$_2$RuO$_4$ sample consists of two B$_{1g}$
phonons.
\end{thebibliography}

\end{document}